\documentclass[prl,twocolumn,nopacs,amsmath,amssymb,superscriptaddress]{revtex4-2}

\usepackage{amsmath}
\usepackage{amssymb}
\usepackage{amsbsy}
\usepackage{graphicx}
\usepackage{xcolor}

\usepackage{mathrsfs}

\usepackage{enumerate}
\usepackage{mathtools}

\usepackage{dcolumn}% Align table columns on decimal point

\newcommand{\moy}[1]{\left\langle #1 \right\rangle}
\newcommand{\dd}[0]{\mathrm{d}}

\newcommand{\erf}[0]{\text{erf}}

\usepackage{soul}

\newcommand{\dt}[2]{\ensuremath{\frac{\dd #1}{\dd #2}}}

\def\e{e}

\DeclareMathOperator{\erfc}{erfc}

\definecolor{darkblue}{rgb}{0,0,0.6}
\definecolor{darkred}{rgb}{0.6,0,0}

\usepackage[colorlinks=true,urlcolor=darkblue,citecolor=darkblue,linkcolor=darkred]{hyperref}

\def\Dt{\tilde{D}}
\def\sgt{\tilde{\sigma}}

\def\qt{\tilde{q}}
\def\pt{\tilde{p}}

\def\rb{\bar\rho}
\def\rbt{\tilde{\bar\rho}}

\def\rt{\tilde{\rho}}

\begin{document}

\title{Tracer diffusion beyond Gaussian behavior:\texorpdfstring{\\}{} explicit results for general single-file systems}

\author{Aur\'elien Grabsch}
\affiliation{Sorbonne Universit\'e, CNRS, Laboratoire de Physique Th\'eorique de la Mati\`ere Condens\'ee (LPTMC), 4 Place Jussieu, 75005 Paris, France}

\author{Olivier B\'enichou}
\affiliation{Sorbonne Universit\'e, CNRS, Laboratoire de Physique Th\'eorique de la Mati\`ere Condens\'ee (LPTMC), 4 Place Jussieu, 75005 Paris, France}

\begin{abstract}
    Single-file systems, in which particles diffuse in narrow channels while not overtaking each other, is a fundamental model for the tracer subdiffusion observed in confined geometries, such as in zeolites or carbon nanotubes.
    Twenty years ago, the mean squared displacement of a tracer was determined at large times, for any diffusive single-file system. Since then, for a general single-file system, even the determination of the fourth cumulant, which probes the deviation from Gaussianity, has remained an open question. Here, we fill this gap and provide an explicit formula for the fourth cumulant of an arbitrary single-file system. Our approach also allows us to quantify the perturbation induced by the tracer on its environment, encoded in the correlation profiles.
    These explicit results constitute a first step towards obtaining a closed equation for the correlation profiles for arbitrary single-file systems.
\end{abstract}

\maketitle

\let\oldaddcontentsline\addcontentsline% Store \addcontentsline
\renewcommand{\addcontentsline}[3]{}% Make \addcontentsline a no-op

\emph{Introduction.---} The investigation of the dynamic properties of interacting particle systems in non-equilibrium settings has been a prominent area of research in last decades~\cite{Spohn:1991,Evans:2005a,Derrida:2007,Chou:2011,Bertini:2015}. Among  them,  single-file  diffusion, where particles diffuse in narrow channels and cannot overtake each other, plays an important role. Such geometrical constraint results in a subdiffusive  behavior of the mean square displacement (MSD) of a tracer particle $\moy{X_T^2} \propto T^{1/2}$~\cite{Harris:1965,Levitt:1973,Arratia:1983}. This theoretical prediction has been verified across various scales, ranging from the diffusion of molecules within zeolites~\cite{Hahn:1996} to the movement of colloids in confined narrow trenches~\cite{Wei:2000,Lin:2005}.

Beyond the scaling behavior of the MSD, the prefactor, which contains the dependence on the mean density $\rb$ of surrounding particles, has first been computed explicitly for specific models: for instance for reflecting Brownian particles~\cite{Harris:1965}, and later for the simple exclusion process (SEP)~\cite{Arratia:1983}.
Twenty years ago, Kollmann extended the result to any single-file system and showed that the MSD of a tracer can be written at large times in terms of macroscopic properties of the system as~\cite{Kollmann:2003}
\begin{equation}
    \label{eq:Kappa2Kollmann}
    \moy{X_T^2} \underset{T \to \infty}{\simeq}
    \frac{\sigma(\rb)}{\rb^2 \sqrt{\pi D(\rb)}} \sqrt{T}
    \:.
\end{equation}
In this expression, $D$ is the collective diffusion coefficient, which controls the relaxation of the density, and $\sigma$ the mobility, which governs the fluctuations of current~\footnote{In the original work of Kollmann~\cite{Kollmann:2003}, the MSD is expressed in terms of $D(\rho)$ and of the static structure factor at vanishing wavenumber $S(\rho)$, which can be related to the mobility and diffusion coefficient as $\sigma(\rho) = 2 \rho D(\rho)S(\rho)$~\cite{Krapivsky:2015a}}. Note that in all these results, as well as throughout this article, annealed (equilibrium) initial conditions have been adopted.

Recently, there has been a growing interest in the characterization of the statistical properties of various observables, and in particular the position of a tracer beyond the MSD (also known as second cumulant)~\cite{Derrida:2009,Derrida:2009a,Krapivsky:2012,Hegde:2014,Krapivsky:2014,Krapivsky:2015a,Sadhu:2015,Imamura:2017,Imamura:2021,Derrida:2019,Derrida:2019b,Mallick:2022,Bettelheim:2022,Bettelheim:2022a,Krajenbrink:2022}. This is typically done by studying higher order cumulants or equivalently the atypical fluctuations using a large deviations framework. These methods give access to finer properties of these observables, beyond the typical Gaussian behavior encoded in the MSD.

More precisely, the higher order cumulants, or large deviations, of the position of the tracer have only been determined for a few specific models. For reflecting Brownian particles the cumulants are known~\cite{Hegde:2014,Krapivsky:2014,Krapivsky:2015a,Sadhu:2015}. For the SEP, all the cumulants have first been determined in the high density limit~\cite{Illien:2013}. At arbitrary density, the computation of the fourth cumulant was first achieved~\cite{Krapivsky:2014,Krapivsky:2015a} and later all the cumulants have been determined~\cite{Imamura:2017,Imamura:2021}.
However, for a general single-file system, even the determination of the fourth cumulant, which probes the deviation from Gaussianity, has remained an open question since the work of Kollmann~\cite{Kollmann:2003}.

Here, we fill this gap and provide an explicit formula for the fourth cumulant for an \textit{arbitrary} single-file system. We stress that, unlike previous results, which were obtained for integrable models (essentially the SEP and those mappable on it~\cite{Rizkallah:2022}) using tools like Bethe ansatz or inverse scattering technique~\cite{Krapivsky:2015a,Imamura:2017,Imamura:2021,Mallick:2022,Bettelheim:2022,Bettelheim:2022a,Krajenbrink:2022}, our expression holds for any model, whether integrable or not. Furthermore, beyond quantifying the deviation from Gaussian behavior, our approach also allows us to quantify the perturbation induced by the tracer on its environment,  encoded in the correlation profiles~\cite{Poncet:2021}. We show that these profiles exhibit a nonanalytic behavior for nonintegrable models.

\emph{Macroscopic fluctuation theory.---} Our starting point to study the position of a tracer in a single-file system relies on the macroscopic fluctuation theory (MFT)~\cite{Bertini:2015}. At large scales (long times and large distances), the MFT gives the probability to observe a fluctuation of the density profile $\rho(x,t)$ of a diffusive system in terms of the two transport coefficients $D(\rho)$ and $\sigma(\rho)$~\cite{Spohn:1983,Derrida:2007,Bertini:2015}, for which explicit expressions have been obtained for several models. For instance, for the SEP, $D(\rho) = 1$ and $\sigma(\rho) = 2 \rho(1-\rho)$. Other paradigmatic models include zero range processes (ZRP)~\cite{Spitzer:1970,Evans:2005a}, the Kipnis-Marchioro-Presutti (KMP) model~\cite{Kipnis:1982}, the Katz–Lebowitz–Spohn (KLS) model~\cite{Katz:1983,Katz:1984}, and models with more realistic pairwise interactions such as Brownian particles with Weeks-Chandler-Anderson (WCA) potential or dipole-dipole interactions, as involved in experimental realisations of colloids confined in 1D~\cite{Wei:2000}. The MFT is a powerful approach, in which all the microscopic details of the model are replaced by the two transport coefficients $D$ and $\sigma$ only. Note however that one typically ends up with nonlinear partial differential equations for the time evolution of the density. Solving these equations is a challenging task of current intense activity which has recently led to important achievements~\cite{Mallick:2022,Bettelheim:2022,Bettelheim:2022a,Krajenbrink:2022,Krajenbrink:2021,Krajenbrink:2022a,Grabsch:2024}.

\begin{figure}
    \centering
    \includegraphics[width=0.9\columnwidth]{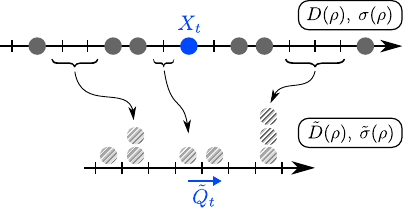}
    \caption{An example of mapping between two single-file systems. The SEP (top) is mapped onto a zero range process (ZRP, below). This well-known mapping holds at the microscopic level: the empty sites of the SEP becomes the particles of the ZRP~\cite{Evans:2005a}, while the position $X_t$ of the tracer in the SEP is mapped onto the integrated current through the origin $\tilde{Q}_t$ in the ZRP. At the macroscopic level, such a mapping holds for any single-file system~\cite{Rizkallah:2022}: the tracer in a system with transport coefficient $D(\rho)$ and $\sigma(\rho)$ is mapped onto the current in a system with $\Dt(\rho)$ and $\sgt(\rho)$ given by~\eqref{eq:MappingTrCoefs}.}
    \label{fig:Mapping}
\end{figure}

MFT has proved to be useful to study a wide range of observables, including even a microscopic observable such as the position $X_T$ of a single tracer, for which different approaches have been devised. (i) The first one relies on expressing $X_T$ as a functional of the density of particles $X_T = X_T[\rho]$~\cite{Krapivsky:2014,Krapivsky:2015a}. This can be done since the tracer effectively ``cuts'' the system into two part, in which the number of particles is conserved due to the noncrossing condition. In practice, this is however tricky due to the emergence of discontinuities in the density profiles at the position of the tracer~\cite{Krapivsky:2014,Krapivsky:2015a}. (ii) A second approach that circumvents this issue consists in introducing a generalised current~\cite{Imamura:2017,Imamura:2021} defined as the number of particles crossing a fictitious moving wall. The tracer is then located at the position where this current vanishes, again due to the noncrossing condition.
(iii) An alternative method, which we apply here, consists in using a mapping between different single-file systems, in which the position $X_T$ of the tracer in the original model is mapped onto (the opposite of) the integrated current $\tilde{Q}_T$ through the origin in a dual model (see Fig.~\ref{fig:Mapping}). More precisely, the current is defined from the density $\tilde\rho(x,t)$ in the dual model as
\begin{equation}
    \label{eq:defQt}
    \tilde{Q}_T = \int_0^\infty \left[ \tilde\rho(x,T) - \tilde\rho(x,0) \right] \dd x
    \:,
\end{equation}
and the transport coefficients $\Dt$ and $\sgt$ of the dual model are written in terms of those of the original model as~\cite{Rizkallah:2022}
\begin{equation}
    \label{eq:MappingTrCoefs}
    \Dt(\rho) = \frac{1}{\rho^2} D \left( \frac{1}{\rho} \right)
    \:,
    \quad
    \sgt(\rho) = \rho \: \sigma \left( \frac{1}{\rho} \right)
    \:.
\end{equation}
The main benefit of this approach is that, with MFT, the study of the current $\tilde{Q}_T$ is generally simpler than that of $X_T$~\cite{Derrida:2009a}. However, this is often at the cost of handling more complex transport coefficients (for instance, the constant $D(\rho) = 1$ of the SEP is mapped onto the non-constant $\Dt(\rho) = 1/\rho^2$). Here, since we aim to study general single-file systems, and thus arbitrary $D$ and $\sigma$, this is not a limitation and we use this latter approach.

The main steps of the computation of the fourth cumulant of $X_T$ for general $D$ and $\sigma$ are as follows (see Supplementary Material (SM) for details~\cite{SM}). First, we use the mapping described above that allows to obtain the cumulants of $X_T$ from those of $\tilde{Q}_T$ in the dual model, with $\Dt$ and $\sgt$ given by~\eqref{eq:MappingTrCoefs}. Explicitly, the cumulant generating functions are related by~\cite{Rizkallah:2022}
\begin{align}
    \hat\psi(\lambda)
    &= \lim_{T \to \infty} \frac{1}{\sqrt{T}} \ln \moy{\e^{\lambda X_T}}
    = \lim_{T \to \infty} \frac{1}{\sqrt{T}} \ln \big\langle \e^{-\lambda \tilde{Q}_T} \big\rangle
    \nonumber
    \\
    &= \kappa_2 \frac{\lambda^2}{2} + \kappa_4 \frac{\lambda^4}{4!} + \cdots
    \:,
\end{align}
with $\kappa_n$ the $n^{\mathrm{th}}$ cumulant of the position of the tracer. Note that the odd order cumulants vanish by symmetry.

Second, we determine the first cumulants of $\tilde{Q}_T$ using the standard MFT formalism~\cite{Derrida:2009a,Bertini:2015}. Explicitly, this requires to solve the MFT equations~\cite{Derrida:2009a}
\begin{align}
  \label{eq:MFT_q}
  \partial_t \qt &= \partial_x[\Dt(\qt) \partial_x \qt]
  - \partial_x[\sgt(\qt)\partial_x \pt]
  \:,
  \\
  \label{eq:MFT_p}
  \partial_t \pt &= - \Dt(\qt) \partial_x^2 \pt
  - \frac{1}{2}  \sgt'(\qt) (\partial_x \pt)^2
  \:,
\end{align}
\begin{equation}
    \pt(x,T) = - \lambda \Theta(x)
    \:,
\end{equation}
\begin{equation}
    \label{eq:InitCond}
    \pt(x,0) = - \lambda \Theta(x)
    + \int_{\rbt}^{\qt(x,0)} \frac{2 \Dt(r)}{\sgt(r)} \dd r
    \:,
\end{equation}
where $\rbt = 1/\rb$ is the mean density in the dual model. The function $\qt(x,t)$ is the typical realisation of the time evolution of the density $\tilde{\rho}(x,t)$ that yields a given value of the current $\tilde{Q}_T$ and fully controls the dynamics at large times $T$. $\pt(x,t)$ is a Lagrange multiplier that ensures the conservation of the number of particles at every point in space and time. The cumulants are then deduced from the solution of these equations by $\dt{\hat\psi}{\lambda} = - \tilde{Q}_T/\sqrt{T}$ where here $\tilde{Q}_T$ is given by~\eqref{eq:defQt} with $\tilde\rho(x,t)$ replaced by its typical fluctuation $\qt(x,t)$.
Third, we expand $\qt$ and $\pt$ in powers of $\lambda$ and solve (\ref{eq:MFT_q}-\ref{eq:InitCond}) order by order, up to order $3$ included to compute $\kappa_4$. The practical resolution requires to solve diffusion equations with source terms of increasing complexity with the order in $\lambda$. Explicit results can be obtained by isolating the dependence of the source terms on $\Dt(\rho)$, $\sgt(\rho)$ and their derivatives, and then relying on a combination of changes of functions and successive integrations by parts.

\emph{Results.---} Lengthy calculations, given in SM~\cite{SM}, finally provide an explicit formula for the fourth cumulant of the position $X_t$ of a tracer for \textit{any} $D(\rho)$ and $\sigma(\rho)$,
\begin{widetext}
\begin{multline}
    \label{eq:Kappa4}
    \kappa_4 =
    \frac{3 \sigma (\rb )^3 \left(\rb  D'(\rb )+D(\rb )\right)}
    {\pi ^{3/2} \rb^6 D(\rb )^{7/2}}
   -\frac{\sigma (\rb )\left(
   \sigma (\rb )\sigma'(\rb ) \left(\rb  D'(\rb )+4 D(\rb )\right)
   + 2 \sigma (\rb )^2 D'(\rb )-\rb  D(\rb ) \sigma'(\rb )^2
   \right)}
   {4 \sqrt{\pi} \rb^5 D(\rb )^{7/2}}
   \\
   +\frac{3 \sigma (\rb )^3  \left(D'(\rb )^2-D(\rb ) D''(\rb )\right)}
   {8 \sqrt{\pi } \rb^4 D(\rb)^{9/2}}
   +\frac{3 \sigma (\rb)^3 \left(2 D(\rb ) D''(\rb )-D'(\rb )^2\right)}
   {8 \pi ^{3/2} \rb ^4 D(\rb )^{9/2}}
   +\frac{\left(3 \sqrt{2}-4\right) \sigma (\rb )^2 \sigma ''(\rb )}
   {8 \sqrt{\pi } \rb ^4 D(\rb )^{5/2}}
   \\
   + \frac{3 \left(\sqrt{2} \pi -2 \sqrt{3}\right) \sigma (\rb)^3 \left(2 D(\rb ) D''(\rb )-3 D'(\rb )^2\right)}
   {16 \pi^{3/2} \rb ^4 D(\rb)^{9/2}}
    \:.
\end{multline}
\end{widetext}
This result constitutes the first step beyond the second cumulant~\eqref{eq:Kappa2Kollmann} for any single-file system and provides a quantitative measure
of the deviation from Gaussian behavior.

Several comments are in order. (i) The expression~\eqref{eq:Kappa4} encompasses all previously known results on fourth cumulants for specific single-file systems, for instance for reflecting Brownian particles~\cite{Hegde:2014,Krapivsky:2014,Krapivsky:2015a,Sadhu:2015}, for the SEP~\cite{Krapivsky:2014,Krapivsky:2015a} and models that can be related to the SEP, such as the KMP model, or the random average process~\cite{Rizkallah:2022}. These previous results were obtained for models that can be mapped, at least at the macroscopic level, to the SEP~\cite{Rizkallah:2022}. For all these models, the last term in~\eqref{eq:Kappa4} vanishes. (ii) More precisely, the last term in Eq.~\eqref{eq:Kappa4} vanishes if and only if $D(\rho) = 1/(a+b \rho)^2$, where $a$ and $b$ are constants. This is the class of diffusion coefficients corresponding to models that can be mapped onto a constant diffusion coefficient (see SM~\cite{SM} for details).
In the general case of a model that cannot be mapped onto a constant $D(\rho)$, as for paradigmatic models like the KLS model or ZRP, or models with more realistic interactions (like Brownian particles with WCA or dipole-dipole interaction) this last term matters. Note that this term is the only one that involves a $\sqrt{3}$. (iii) Finally, the result~\eqref{eq:Kappa4}, also gives the fourth cumulant of the current $\tilde{Q}_T$, in the dual model with $\Dt$ and $\sgt$. Writing this expression in terms of these dual transport coefficients thanks to~\eqref{eq:MappingTrCoefs}, gives this fourth cumulant of $\tilde{Q}_T$ for a general single-file system (see Eq.~(S60) of the SM).

\emph{Beyond the cumulants: correlation profiles.---} On top of the cumulants, our approach gives access to the response of the bath of surrounding particles to the perturbation induced by the displacement of the tracer. This response is described by the bath-tracer correlation profile introduced in~\cite{Poncet:2021}, defined as
\begin{align}
    \nonumber
    w(x,T)
    &\equiv
    \frac{\moy{ \rho(X_T+x,T) \e^{\lambda X_T}}}
    {\moy{\e^{\lambda X_T}}}
    \\
    &= \sum_{n=0}^\infty \frac{\lambda^n}{n!} \moy{\rho(X_T+x,T) X_T^n}_c
    \label{eq:CorrProf}
    \:,
\end{align}
which generates all the connected correlation functions $\moy{\rho(X_T+x,T) X_T^n}_c$ between the density field and the displacement of the tracer.
At large times $T$, these profiles display a diffusive scaling behavior $w(x,T) \simeq \Phi(z = x/\sqrt{T})$~\cite{Poncet:2021,Grabsch:2022,Grabsch:2023}. The scaling function $\Phi$, which thus contains the full spatial structure of the bath-tracer correlations in the long time limit, has been determined explicitly for the SEP and for models that can be related to it~\cite{Grabsch:2022,Grabsch:2023,Rizkallah:2022}.
Here, for arbitrary $D(\rho)$ and $\sigma(\rho)$, $\Phi$ is derived from the solution of the MFT equation $\tilde{q}(x,T)$ at final time (in the dual model with $\Dt$ and $\sgt$) and mapped back to the original model with $D$ and $\sigma$. The details of this mapping and the expressions of the correlation profiles up to order 3 are given explicitly in SM~\cite{SM}, Eqs.~(S67-S69).

In parallel of this explicit calculation, an important question concerns the existence of a closed equation satisfied by $\Phi$. Indeed, in the case of the SEP, these profiles have been shown to satisfy a simple exact closed equation~\cite{Grabsch:2022,Grabsch:2023}. This result has allowed the determination of all the correlation profiles~\eqref{eq:CorrProf}.
Since the publication of this equation~\cite{Grabsch:2022}, several works have obtained exact results for different observables for specific models of single-file systems~\cite{Mallick:2022,Bettelheim:2022,Bettelheim:2022a,Krajenbrink:2022}, which can all be recast into a similar closed equation, making it a promising tool to investigate various questions in single-file diffusion and beyond.

We investigate the possibility to obtain such an equation for $\Phi$ by following the approach of~\cite{Grabsch:2022,Grabsch:2023}. It is shown in SM~\cite{SM} that in fact,
\begin{widetext}
\begin{multline}
    \label{eq:BulkEq}
    \partial_z(D(\Phi)\partial_z \Phi)
    + \frac{1}{2}(z+\xi) \partial_z \Phi
    = \frac{\lambda \sigma''(\rb)}{4\rb} \int_0^\infty \Phi'(z+u)\Phi''(-u) \dd u
    \\
    + \left( \frac{\lambda  \sigma (\rb ) D'(\rb )}{8 \sqrt{\pi } \rb  D(\rb )^{3/2}}
    -\frac{\lambda ^2 \sigma (\rb ) D'(\rb ) \left(\rb  \sigma '(\rb )-2 \sigma (\rb)\right)}{32 \sqrt{\pi } \rb ^3 D(\rb )^{5/2}}
    +\frac{\lambda ^2 \sigma (\rb )^2 D'(\rb )^2}{64 \sqrt{\pi } \rb ^2 D(\rb)^{7/2}}
    \right) \Phi'(z)
    \\
    - \frac{\lambda ^3 \sigma (\rb )^3 \left(2 D(\rb ) D''(\rb )-3 D'(\rb )^2\right)}{512
   \rb ^3 D(\rb )^5}
   \left(
   y \e^{-\frac{y^2}{2}} \sqrt{\frac{2}{\pi}} \erfc \left( \frac{y}{\sqrt{2}} \right)
   +\frac{2}{\pi^{3/2}} \partial_y \int_0^1 \frac{\dd t}{\sqrt{1+2t}}
    \e^{- \frac{(1+t) y^2}{(1-t)(1+2t)}}
   \right)
   + \mathcal{O}(\lambda^4)
   \:,
\end{multline}
\end{widetext}
where $\xi = \dt{\hat\psi}{\lambda}$, and we have denoted $y = \frac{z}{2\sqrt{D(\rb)}}$ to simplify the notations. In the case of the SEP, corresponding to constant $D$, only the first term on the r.h.s. of Eq.~\eqref{eq:BulkEq} remains. We have written  this term as a convolution, instead of its explicit expression, because it was the key step in~\cite{Grabsch:2022,Grabsch:2023} that allowed to find a closed form for the equation. Similarly, we have realized that the second term in~\eqref{eq:BulkEq} can be expressed in terms of $\Phi'$ only.
The only remaining task to obtain a closed equation is to rewrite the last term in~\eqref{eq:BulkEq} in terms of $\Phi$.
Anyhow, Eq.~\eqref{eq:BulkEq} constitutes a first step towards obtaining a closed equation for $\Phi$ for arbitrary $D(\rho)$ and $\sigma(\rho)$.

On top of its intrinsic interest, Eq.~\eqref{eq:BulkEq} allows us, as we now discuss, to provide (i) a signature of the nonintegrable nature of general single-file models and (ii) a shortcut to obtain the cumulants of $X_t$.

\emph{Relation with integrability.---}
First, it can be shown that the last term in~\eqref{eq:BulkEq} is directly associated to the $\sqrt{3}$ in the expression of $\kappa_4$~\eqref{eq:Kappa4}, as discussed above.
In particular, both vanish for the specific choice $D(\rho) = 1/(a + b \rho)^2$, which is the class of diffusion coefficients for which the nonlinear heat equation is integrable~\cite{Liu:2016}.
Second, this term displays a nonanalytic behavior with respect to the distance to the tracer, with a logarithmic singularity $\sim y \ln y$ as $y \to 0$. It shows that this term introduces a completely new class of functions, compared to the case of the SEP (and related models) in which only analytic functions were present~\cite{Grabsch:2022,Grabsch:2023}.
Note that such behavior was also observed in the correlation profile of a \textit{driven} tracer in the SEP~\cite{Grabsch:2023b},  a model which is expected to be not integrable. All these points indicate that the presence of the last term in~\eqref{eq:BulkEq} is a signature of the nonintegrability of a single-file model with arbitrary $D(\rho)$ and $\sigma(\rho)$.

\emph{A conjecture for a shortcut to the cumulants.---} First of all, we remind that in the case of the SEP, boundary conditions for $\Phi(0^\pm)$ and $\Phi'(0^\pm)$ have been obtained from microscopic considerations~\cite{Poncet:2021,Grabsch:2022,Grabsch:2023}. These relations are very useful, since together with the bulk equation~\eqref{eq:BulkEq} written in the specific case of the SEP, they allow to fully determine the profiles and the cumulants without solving the MFT equations~(\ref{eq:MFT_q}-\ref{eq:InitCond}). Several of these relations have recently been extended to any single-file system, and take a simple physical form~\cite{Grabsch:2024}
\begin{equation}
    \label{eq:BoundCondPhi}
    P(\Phi(0^+)) - P(\Phi(0^-)) = \lambda
    \:,
    \quad
    \left[ \partial_z \mu(\Phi) \right]_{0^-}^{0^+}
    = 0
    \:,
\end{equation}
where $P(\rho)$ is the pressure, and $\mu(\rho)$ the chemical potential, given by $P'(\rho) = \rho \mu'(\rho)$ and $\mu'(\rho) = 2 D(\rho)/ \sigma (\rho)$. We have used the notation $[f]_a^b = f(b) - f(a)$. The remaining boundary condition,  obtained for the SEP in \cite{Poncet:2021,Grabsch:2022,Grabsch:2023}, which has not yet been generalized to an arbitrary single-file system~\cite{Grabsch:2024}, is a key relation allowing to obtain $\hat\psi$ directly from $\Phi(0^\pm)$ and $\Phi'(0^\pm)$ (which are fully determined by the bulk equation~\eqref{eq:BulkEq} and the boundary conditions~\eqref{eq:BoundCondPhi} completed by $\Phi(\pm \infty) = \rb$), instead of computing the integral~\eqref{eq:defQt}, which is usually a difficult task. We conjecture that, for arbitrary $D$ and $\sigma$, this last relation takes the form
\begin{equation}
    \label{eq:ConjecPsi}
    \hat\psi = - 2 \left. \partial_z \mu(\Phi) \right|_{z=0}
    \int_{\Phi(0^-)}^{\Phi(0^+)} D(r) \dd r
    \:.
\end{equation}
This conjecture is supported by the following points. (i) For $D(\rho)=1$ and $\sigma(\rho) = 2 \rho(1-\rho)$, it reduces to the expression obtained for the SEP~\cite{Poncet:2021,Grabsch:2022,Grabsch:2023}. (ii) Furthermore, from our above results on the profiles $\Phi$ and the fourth cumulant $\kappa_4$, we can check that this relation holds up to order $4$ in $\lambda$ included. (iii) Finally, Eq.~\eqref{eq:ConjecPsi} is invariant under the duality mapping~\eqref{eq:MappingTrCoefs} (see SM~\cite{SM}).

\emph{Conclusion.---} We have considered tracer diffusion (as well as the current of particles) in general single-file systems at large times. We have determined an explicit expression for the fourth cumulant of $X_t$, which constitutes the first extension of the result of Kollmann on the second cumulant~\cite{Kollmann:2003}, for any $D(\rho)$ and $\sigma(\rho)$ and provides a quantitative measure of the deviation from Gaussian behavior.
On top of the cumulants of $X_t$, we have obtained the response of the bath of surrounding particles to the displacement of the tracer by determining the full spatial structure of the bath-tracer correlation profiles (up to order $4$).
These explicit results, which hold for  any system, allowed us to pinpoint the effect of nonintegrability, both on the cumulants and the correlation profiles.
This work constitutes a first step towards obtaining a closed equation for the correlation profiles for arbitrary $D(\rho)$ and $\sigma(\rho)$.

% \emph{Acknowledgements.---}

\bibliographystyle{apsrev4-2}
% \bibliography{bibSF}


\clearpage
% \newpage
\widetext

\let\addcontentsline\oldaddcontentsline% Restore \addcontentsline

\begin{center}
  \begin{large}

    \textbf{
     Supplementary Material for\texorpdfstring{\\}{} Tracer diffusion beyond Gaussian behavior:\texorpdfstring{\\}{} explicit results for general single-file systems
   }
  \end{large}
   \bigskip

    Aurélien Grabsch and Olivier Bénichou
\end{center}

\setcounter{equation}{0}
\makeatletter

\renewcommand{\theequation}{S\arabic{equation}}
\renewcommand{\thefigure}{S\arabic{figure}}

\renewcommand{\bibnumfmt}[1]{[S#1]}
\renewcommand{\citenumfont}[1]{S#1}

\setcounter{secnumdepth}{3}

\tableofcontents

\section{Macroscopic fluctuation theory for a tracer}

In this Section, we summarise the main tools needed to study the position of a tracer at large times in a single-file system.

\subsection{The transport coefficients}

At large times and large distances, a single-file system can be described by a density field $\rho(x,t)$ which obeys a stochastic diffusion equation~\cite{Spohn:1983SM}
\begin{equation}
    \label{eq:StochDiff}
    \partial_t \rho = \partial_x \left[
    D(\rho) \partial_x \rho
    + \sqrt{\sigma(\rho)} \eta
    \right]
    \:,
\end{equation}
where $\eta$ is a Gaussian white noise in space and time, with $\moy{\eta(x,t) \eta(x',t')} = \delta(x-x') \delta(t-t')$. This equation involves only two transport coefficients: the diffusion coefficient $D(\rho)$ and the mobility $\sigma(\rho)$. These quantities were first defined for a lattice gas~\cite{Spohn:1983SM}, but they can be defined more intuitively for any single-file system by considering a finite system of length $L$ between two reservoirs at densities $\rho_{\mathrm{L}}$ and $\rho_{\mathrm{R}}$~\cite{Derrida:2007SM}. Consider the total number $Q_t$ transferred from the left reservoir to the right one. The diffusion coefficient measures the average current in the presence of a small difference of density,
\begin{equation}
    \lim_{t \to \infty} \frac{\moy{Q_t}}{t} =
    \frac{D(\rho)}{L} (\rho_{\mathrm{L}} - \rho_{\mathrm{R}})
    \:,
    \quad
    \text{for}
    \quad
    \rho_{\mathrm{R}} - \rho_{\mathrm{L}} \ll
    \rho \equiv \frac{\rho_{\mathrm{R}} + \rho_{\mathrm{L}}}{2}
    \:.
\end{equation}
The mobility measures the fluctuations of current at a given density
\begin{equation}
    \lim_{t \to \infty} \frac{\moy{Q_t}^2}{t} =
    \frac{\sigma(\rho)}{L}
    \:,
    \quad
    \text{for}
    \quad
    \rho = \rho_{\mathrm{R}} = \rho_{\mathrm{L}}
    \:.
\end{equation}
These two coefficients fully describe the single-file system at the macroscopic level.

\subsection{Mapping onto a dual problem of current}

Our computations rely on a mapping between single-file systems. The density $\rho(x,t)$ of the original single-file system can be mapped onto another density
\begin{equation}
    \label{eq:MappingDens}
    \tilde\rho(y,t) = \frac{1}{\rho(x(y,t),t)}
    \:,
    \quad
    x(y,t) = \int_0^y \rt(y',t) \dd y' + X_t
    \:,
  \end{equation}
where we denote $X_t$ the position of a tracer, initially at the origin $X_0 = 0$.
Under this transformation, the density $\tilde\rho$ describes another single-file system~\cite{Rizkallah:2022SM},
\begin{equation}
    \partial_t  \tilde\rho = \partial_y \left[
    \Dt(\rho) \partial_y \tilde\rho
    + \sqrt{\tilde\sigma(\rho)} \eta
    \right]
    \:,
\end{equation}
with new transport coefficients
\begin{equation}
    \label{eq:MappingTrCoefsSM}
    \Dt(\rho) = \frac{1}{\rho^2} D \left( \frac{1}{\rho} \right)
    \:,
    \quad
    \sgt(\rho) = \rho \: \sigma \left( \frac{1}{\rho} \right)
    \:.
\end{equation}
This is a generalisation of previously known mappings between specific models of single-file systems, such as between the SEP and the zero range process~\cite{Evans:2000SM,Evans:2005SM}, and between the
random average process (RAP) and the Kipnis Marchioro Presutti (KMP) model~\cite{Kundu:2016SM}.
The idea behind this transformation is that a given single-file system can be viewed in two equivalent ways: (i) either by looking at the positions of particles, and thus their density $\rho(x,t)$ (ii) or by looking at the distances between the particles. In this second point of view, the variable $y$ represents the ''label'' of the particle located at position $x$ at time $t$, while $\tilde\rho(y,t)$ is the distance between this particle and the next one. Although equivalent, these two points of view give rise to two different models of single-file systems, with different transport coefficients related by~\eqref{eq:MappingTrCoefsSM}.

\bigskip

In the original system, at $t=0$ we start in an equilibrium configuration at mean density $\rb$. From the mapping~\eqref{eq:MappingDens}, this becomes an equilibrium at mean density \begin{equation}
\label{eq:TrMeanDens}
    \rbt = \frac{1}{\rb}
    \:.
\end{equation}

\bigskip

The position $X_t$ of the tracer in the original system can equivalently be written in terms of the variation of the distance between the particles, and thus reads in terms of the dual model~\cite{Rizkallah:2022SM},
\begin{equation}
    \label{eq:RelXtQt}
    X_t = - \int_0^\infty \left[
    \tilde\rho(y,t) - \tilde\rho(y,0)
    \right]
    \dd y
    \equiv
    - \tilde{Q}_t
    \:,
\end{equation}
where the r.h.s. is actually the (opposite of) the integrated current through the origin in the dual model (equivalently given by the variation of the total density at the right of the origin). Therefore, the problem of studying the dynamics of a tracer in a given single-file model reduces to the study of the current in the dual single-file system.

\subsection{MFT equations for the study of the current}
\label{sec:MFT}

The MFT can be applied to study the statistical properties of the current $\tilde{Q}_T$ at a large time $T$ in a single-file system with arbitrary transport coefficients $\Dt$ and $\sgt$~\cite{Derrida:2009aSM}. We sketch here the main steps of the derivation of the MFT equations. The moment generating function of the position $X_T$ of the tracer can be written as
\begin{equation}
    \moy{\e^{\lambda X_T}} = \moy{\e^{-\lambda \tilde{Q}_T}}
    = \int \mathcal{D} \tilde\rho(x,t) \mathcal{D} \tilde{H}(x,t)
    \int \mathcal{D} \tilde\rho(x,0) \:
    \e^{-\lambda Q_T[\tilde\rho] - S[\tilde{\rho},\tilde{H}] - F[\tilde{\rho}(x,0)]}
    \:,
\end{equation}
where $S$ is the MFT action ($\tilde{H}$ is a Lagrange multiplier that enforces the local conservation of particles)
\begin{equation}
    S[\tilde\rho,\tilde{H}] =
    \int_{-\infty}^\infty \dd x \int_0^T \dd t \left[
    \tilde{H} \partial_t \tilde\rho
    + \Dt(\tilde\rho) \partial_x \tilde\rho \partial_x \tilde{H}
    - \frac{\sgt(\tilde\rho)}{2} (\partial_x \tilde{H})^2
    \right]
    \:,
\end{equation}
$F$ gives the distribution of the initial condition picked from an equilibrium density $\rbt$,
\begin{equation}
    F[\tilde\rho(x,0)] = \int_{-\infty}^\infty \dd x
    \int_{\rbt}^{\tilde\rho(x,0)} \dd r
    \left[ \tilde\rho(x,0) - r \right] \frac{2 D(r)}{\sigma(r)}
    \:,
\end{equation}
and the functional $Q_T$ gives the integrated current associated to the time evolution $\tilde\rho(x,t)$,
\begin{equation}
    Q_T[\tilde\rho] = \int_0^\infty \left[
    \tilde\rho(x,T) - \tilde{\rho}(x,0)
    \right]
    \:.
\end{equation}
Rescaling $x$ by $\sqrt{T}$ and $t$ by $T$ we obtain (with a slight abuse of notation, we keep the same names for the new fields of rescaled variables),
\begin{equation}
    \label{eq:MomGenFctMFT}
    \moy{\e^{\lambda X_T}} = \moy{\e^{-\lambda \tilde{Q}_T}}
    = \int \mathcal{D} \tilde\rho(x,t) \mathcal{D} \tilde{H}(x,t)
    \int \mathcal{D} \tilde\rho(x,0) \:
    \e^{- \sqrt{T}(\lambda Q[\tilde\rho] + S[\tilde{\rho},\tilde{H}] + F[\tilde{\rho}(x,0)])}
    \:,
\end{equation}
with now the time $t$ belonging to $[0,1]$, i.e.,
\begin{equation}
    S[\tilde\rho,\tilde{H}] =
    \int_{-\infty}^\infty \dd x \int_0^1 \dd t \left[
    \tilde{H} \partial_t \tilde\rho
    + \Dt(\tilde\rho) \partial_x \tilde\rho \partial_x \tilde{H}
    - \frac{\sgt(\tilde\rho)}{2} (\partial_x \tilde{H})^2
    \right]
    \:,
    \quad
    Q[\tilde\rho] = \int_0^\infty \left[
    \tilde\rho(x,1) - \tilde{\rho}(x,0)
    \right]
    \:.
\end{equation}
Thanks to the factor $\sqrt{T}$ in the exponential in~\eqref{eq:MomGenFctMFT}, the functional integrals can be evaluated by a saddle point method. Let us denote $(\qt,\pt)$ the fields $(\tilde\rho,\tilde{H})$ which minimize $S + F + \lambda Q$. They satisfy the MFT equations~\cite{Derrida:2009aSM}
\begin{align}
  \label{eq:MFT_qSM}
  \partial_t \qt &= \partial_x[\Dt(\qt) \partial_x \qt]
  - \partial_x[\sgt(\qt)\partial_x \pt]
  \:,
  \\
  \label{eq:MFT_pSM}
  \partial_t \pt &= - \Dt(\qt) \partial_x^2 \pt
  - \frac{1}{2}  \sgt'(\qt) (\partial_x \pt)^2
  \:,
\end{align}
with the final condition for $\pt$
\begin{equation}
    \pt(x,1) = - \lambda \Theta(x)
    \:,
\end{equation}
and the initial condition for $\qt$
\begin{equation}
    \label{eq:InitTimeSM}
    \pt(x,0) = - \lambda \Theta(x)
    + \int_{\rbt}^{\qt(x,0)} \frac{2 \Dt(r)}{\sgt(r)} \dd r
    \:.
\end{equation}
The cumulant generating function is given by
\begin{equation}
\hat\psi(\lambda) \equiv \lim_{T \to \infty} \frac{1}{\sqrt{T}}
    \ln \moy{\e^{\lambda X_T}}
    = \lim_{T \to \infty} \frac{1}{\sqrt{T}} \ln \moy{\e^{-\lambda Q_T}}
    =  - \left( \lambda Q[\qt] + S[\qt,\pt] + F[\qt(x,0)] \right)
    \:.
\end{equation}
In practice, it is simpler to compute the derivative of the cumulant generating function,
\begin{equation}
    \label{eq:CumulGenFctFromQ0}
    \dt{}{\lambda} \hat\psi
    = - Q[\tilde{q}]
\end{equation}
since $\qt$ and $\pt$ minimize the action. These are the equations~(6-8) in the main text (up to the rescaling by $\sqrt{T}$). Finally, in~\cite{Derrida:2009aSM}, it was noted that the MFT equations~(\ref{eq:MFT_qSM}-\ref{eq:InitTimeSM}) have a time-reversal symmetry:
\begin{equation}
\label{eq:TimeRevSym}
    \left. \qt(x,1-t) \right|_{\lambda \to - \lambda}
    = \qt(x,t)
    \:.
\end{equation}
This implies that the cumulant generating function~\eqref{eq:CumulGenFctFromQ0} can be expressed in terms of the profile at final time only:
\begin{equation}
    \label{eq:CumulGenFctFromQ}
    \dt{}{\lambda} \hat\psi
    = - \int_0^\infty \left[
    \qt(x,1) - \left. \qt(x,1)  \right|_{\lambda \to - \lambda}
    \right] \dd x
    \:.
\end{equation}
This will be useful to compute the cumulants below.

\section{Perturbative solution of the MFT equations}
\label{sec:PertSol}

We look for a perturbative solution by expanding the functions $\pt$ and $\qt$ in powers of $-\lambda$,
\begin{equation}
    \label{eq:Exppqlambda}
    \pt = -\lambda \pt_1 + \lambda^2 \pt_2 - \lambda^3 \pt_3 + \mathcal{O}(\lambda^4)
    \:,
    \qquad
    \qt = \rbt - \lambda \qt_1 + \lambda^2 \qt_2 - \lambda^3 \qt_3 + \mathcal{O}(\lambda^4)
    \:.
\end{equation}

For simplicity, all the results will be expressed in terms of the rescaled variable
\begin{equation}
    y \equiv \frac{x}{\sqrt{4 \Dt(\rbt)}}
    \:.
\end{equation}

\subsection{First order}

At first order in $\lambda$, the solution of the MFT equations is given by
\begin{equation}
    \label{eq:p1}
    \pt_1(x,t) = \frac{1}{2} \erfc \left( -\frac{y}{\sqrt{1-t}} \right)
    \:,
\end{equation}
\begin{equation}
    \label{eq:q1}
    \qt_1(x,t) = \frac{\sgt(\rbt)}{4 \Dt(\rbt)} \left[
         \erfc \left( \frac{y}{\sqrt{t}} \right)
         -  \erfc \left( \frac{y}{\sqrt{1-t}} \right)
    \right]
    \:.
\end{equation}

\subsection{Second order}

At second order, the solution of the MFT equations reads
\begin{equation}
    \label{eq:p2}
    \pt_2(x,t) = \frac{\sgt(\rbt)}{16 \Dt(\rbt)}
    \erfc \left( \frac{y}{\sqrt{1-t}} \right)
    \erfc \left( -\frac{y}{\sqrt{1-t}} \right)
    + \frac{\Dt'(\rbt) \sgt(\rbt)}{\Dt(\rbt)^2}
    Q_2(y,1-t)
    \:,
\end{equation}
\begin{multline}
    \label{eq:q2}
    \qt_2(x,t) = \frac{\sgt(\rbt)}{2 \Dt(\rbt)} \pt_2(x,t)
    + \frac{1}{2} \left( \frac{\sgt'(\rbt)}{\sgt(\rbt)} - \frac{\Dt'(\rbt)}{\Dt(\rbt)} \right) \qt_1(x,t)^2
    + \frac{\sgt'(\rbt) \sgt(\rbt)}{32 \Dt(\rbt)^2}
    \erfc \left( \frac{y}{\sqrt{t}} \right)
    \erfc \left( -\frac{y}{\sqrt{t}} \right)
    \\
    + \frac{\Dt'(\rbt) \sgt(\rbt)^2}{2 \Dt(\rbt)^3} Q_2(y,t)
    \:,
\end{multline}
where $Q_2$ is the solution of the equation
\begin{equation}
    \partial_t Q_2 - \frac{1}{4} \partial_y^2 Q_2 =
    \frac{y \: \e^{-\frac{y^2}{t}} }{8 \sqrt{\pi} t^{3/2}}
    \left[
    \erfc\left( \frac{y}{\sqrt{t}} \right)
    -\erfc\left( \frac{y}{\sqrt{1-t}} \right)
    \right]
    \:,
    \quad
    Q_2(y,0) = 0
    \:.
\end{equation}
Explicitly, this function takes the form
\begin{multline}
    Q_2(y,t) = - \frac{\e^{-\frac{2 y^2}{t}}}{2\pi}
    - \frac{\e^{-\frac{y^2}{t}}}{8\pi \sqrt{t}}
    + \frac{\e^{- \frac{y^2}{t} - \frac{y^2}{1-t}}}{8\pi} \sqrt{\frac{1-t}{t}}
    \\
    + \frac{y \: \e^{-\frac{y^2}{t}} }{8 \sqrt{\pi \: t} }
    \left[
    \erfc\left( \frac{y}{\sqrt{t}} \right)
    -\erfc\left( \frac{y}{\sqrt{1-t}} \right)
    \right]
    + \frac{1}{4} \mathrm{T} \left( y \sqrt{\frac{2}{t}}, \sqrt{\frac{t}{1-t}} \right)
    \:,
\end{multline}
where $T$ is the Owen-T function, defined by~\cite{Owen:1980SM}
\begin{equation}
  \label{eq:defOwenT}
  T(h,a) = \frac{1}{2\pi}
  \int_0^a \frac{\e^{-\frac{h^2}{2}(1+x^2)}}{1+x^2} \dd x
  \:.
\end{equation}

\subsection{Third order}

Since we only need $\qt_3$ at final time $t=1$, we do not need to determine $\pt_3$. We can obtain equations for $\qt_3$ only by writing it as
\begin{multline}
    \label{eq:q3}
    \qt_3(x,t) = \frac{\sgt(\rbt)}{2 \Dt(\rbt)} \pt_3(x,t)
    +\left(\frac{\sgt'(\rbt)}{\sgt(\rbt)} - \frac{\Dt'(\rbt)}{\Dt(\rbt)} \right)
    \qt_1(x,t) \qt_2(x,t)
    + \frac{\sgt'(\rbt)^2 \sgt(\rbt)}{192 \Dt(\rbt)^3}
    \erf \left( \frac{y}{\sqrt{t}} \right)
    \erfc \left( \frac{y}{\sqrt{t}} \right)
    \erfc \left( -\frac{y}{\sqrt{t}} \right)
    \\
    - \frac{\qt_1(x,t)^3}{6 \Dt(\rbt) \sgt(\rbt)^2}
    \left[
    2 \Dt(\rbt) \sgt'(\rbt)^2 + \sgt(\rbt)^2 \Dt''(\rbt)
    - \sgt(\rbt) \left(
    \Dt(\rbt) \sgt''(\rbt) + 2 \Dt'(\rbt) \sgt'(\rbt)
    \right)
    \right]
    + \frac{\sgt''(\rbt) \sgt(\rbt)}{\Dt(\rbt)}
    Q_{3;\sgt''} (y,t)
    \\
    + \frac{\Dt''(\rbt) \sgt(\rbt)^3}{\Dt(\rbt)^4}
    Q_{3;\Dt''}(y,t)
    + \frac{\Dt'(\rbt) \sgt'(\rbt) \sgt(\rbt)^2}{\Dt(\rbt)^4}
    Q_{3;\Dt',\sgt'}(y,t)
    + \frac{\Dt'(\rbt)^2 \sgt(\rbt)^3}{\Dt(\rbt)^5}
    Q_{3;(\Dt')^2}(y,t)
    \:,
\end{multline}
where all the dependence on $\Dt$ and $\sgt$ has been extracted.
The functions $Q_{3;\sgt''}$, $Q_{3;\Dt''}(y,t)$, $Q_{3;\Dt',\sgt'}$ and $Q_{3;(\Dt')^2}$ are solutions of different inhomogeneous diffusion equations, with source terms that are explicit thanks to the expressions~\eqref{eq:p1},\eqref{eq:q1},\eqref{eq:p2},\eqref{eq:q2} obtained at previous orders.

\subsubsection{The term in \texorpdfstring{$\sgt''$}{sigma}}

The function $Q_{3;\sgt''} (y,t)$ is solution of
\begin{equation}
    \label{eq:Q3sigma}
    \partial_t Q_{3;\sgt''} (y,t)
    - \frac{1}{4} \partial_y^2 Q_{3;\sgt''} (y,t)
    =
    \frac{\e^{-\frac{2y^2}{t}}}{64 \pi t}
    \left[
    \erfc \left( \frac{y}{\sqrt{t}} \right)
    - \erfc \left( \frac{y}{\sqrt{1-t}} \right)
    \right]
    \:,
    \quad
    Q_{3;\sgt''}(y,0) = 0
    \:.
\end{equation}
The solution at final time $Q_{3;\sgt''} (y,1)$ has been determined in~\cite{Grabsch:2022SM,Grabsch:2023SM}.
We recall here the steps leading to this result, as a similar approach will be used for the other terms below. First, we notice that the first term in the r.h.s. of~\eqref{eq:Q3sigma} is a function of $y/\sqrt{t}$ only. We can thus look for a solution in the form of a scaling function $g(y/\sqrt{t})$,
\begin{equation}
    \partial_t g \left( \frac{y}{\sqrt{t}} \right)
    - \frac{1}{4} \partial_y^2 g \left( \frac{y}{\sqrt{t}} \right)
    = \frac{\e^{-\frac{2y^2}{t}}}{64 \pi t}
    \erfc \left( \frac{y}{\sqrt{t}} \right)
    \quad \Rightarrow \quad
    g''(z) +2 z g'(z) = - \frac{\e^{-2z^2}}{16 \pi} \erfc(z)
    \:,
\end{equation}
from which we deduce
\begin{equation}
    g(z) = a + b \erfc(z) - \frac{1}{384} \erfc(z)^3
    \:.
\end{equation}
Imposing $g(\pm \infty) = 0$ so that $g(y/\sqrt{t}) \to 0$ for $t\to 0$, this gives
\begin{equation}
    g(z) = \frac{1}{96} \erfc(z) - \frac{1}{384} \erfc(z)^3
    \:.
\end{equation}

We now perform the change of functions
\begin{equation}
    Q_{3;\sgt''} (y,t) =
    \frac{1}{96} \erfc \left( \frac{y}{\sqrt{t}} \right)
    - \frac{1}{384} \erfc \left( \frac{y}{\sqrt{t}} \right)^3
    + \tilde{Q}_{3;\sgt''} (y,t)
    \:,
\end{equation}
where $\tilde{Q}_{3;\sgt''}$ now verifies
\begin{equation}
    \partial_t \tilde{Q}_{3;\sgt''} (y,t)
    - \frac{1}{4} \partial_y^2 \tilde{Q}_{3;\sgt''} (y,t)
    =
    -\frac{\e^{-\frac{2y^2}{t}}}{64 \pi t}
    \erfc \left( \frac{y}{\sqrt{1-t}} \right)
    \:,
    \quad
    \tilde{Q}_{3;\sgt''}(y,0) = 0
    \:.
\end{equation}
The solution can be written as a double convolution with the heat kernel. Since we are only interested in the solution at final time $t=1$, this gives,
\begin{equation}
    \tilde{Q}_{3;\sgt''} (y,1)
    = -\int_{-\infty}^\infty \dd z \int_0^1 \dd t
    \frac{\e^{-\frac{2z^2}{t}}}{64 \pi t}
    \erfc \left( \frac{z}{\sqrt{1-t}} \right)
    \frac{\e^{-\frac{(z-y)^2}{1-t}}}{\sqrt{\pi(1-t)}}
    \:.
\end{equation}
The integral over $z$ can be computed using the table~\cite{Owen:1980SM}, and gives
\begin{equation}
    \label{eq:IntegReprQ3tSig}
    \tilde{Q}_{3;\sgt''} (y,1)
    = - \int_0^1 \frac{\dd t}{64 \pi} \frac{\e^{- \frac{2y^2}{2-t}}}{\sqrt{t(2-t)}}
    \erfc \left( \frac{t y}{\sqrt{2(2-3t+t^2)}} \right)
    \:,
\end{equation}
from which we deduce
\begin{equation}
    \label{eq:ValuesQ30}
    \tilde{Q}_{3;\sgt''} (0,1) = - \frac{1}{128}
    \:,
    \quad
    \left. \partial_y \tilde{Q}_{3;\sgt''} (y,1) \right|_{y=0}
    = \frac{\sqrt{2}-1}{32 \sqrt{2 \pi}}
    \:.
\end{equation}
We have not been able to directly evaluate~\eqref{eq:IntegReprQ3tSig}, but we can compute
\begin{equation}
    \partial_y^2 \tilde{Q}_{3;\sgt''} (y,1) + 2 y \partial_y  \tilde{Q}_{3;\sgt''} (y,1)
    = \int_0^1 \frac{\dd t}{32 \sqrt{2} \pi^{3/2} t^{3/2}} y \sqrt{1-t} \e^{-\frac{1+t}{2t}y^2}
    \:,
\end{equation}
after integration by parts. This last integral can be computed, and yields,
\begin{equation}
    \partial_y^2 \tilde{Q}_{3;\sgt''} (y,1) + 2 y \partial_y  \tilde{Q}_{3;\sgt''} (y,1)
    = \frac{\e^{-y^2}}{32 \pi}
    - \frac{y \: \e^{-\frac{y^2}{2}}}{32 \sqrt{2\pi}} \erfc \left( \frac{y}{\sqrt{2}} \right)
    \:.
\end{equation}
Solving this differential equation with the initial conditions~\eqref{eq:ValuesQ30}, we get,
\begin{equation}
    \tilde{Q}_{3;\sgt''} (y,1) = \frac{1}{128} \erfc\left( \frac{y}{\sqrt{2}} \right)^2
    - \frac{1}{64} \erfc(y)
    \:.
\end{equation}
Finally, this gives the result of~\cite{Grabsch:2022SM,Grabsch:2023SM},
% by manipulating the integral representations of the solution of the heat equation~\eqref{eq:Q3sigma}. It reads
\begin{equation}
    Q_{3;\sgt''} (y,1) = \frac{1}{128} \erfc \left( \frac{y}{\sqrt{2}} \right)^2
    - \frac{1}{192} \erfc(y)
    - \frac{1}{384} \erfc(y)^3
    \quad
    \text{for}
    \quad
    y > 0
    \:.
\end{equation}

\subsubsection{The term in \texorpdfstring{$\Dt''$}{D}}

The function $Q_{3;\Dt''} (y,t)$ is solution of
\begin{equation}
    \label{eq:Q3D2}
    \partial_t Q_{3;\Dt''} (y,t)
    - \frac{1}{4} \partial_y^2 Q_{3;\Dt''} (y,t)
    =
    \frac{y \: \e^{-\frac{y^2}{t}}}{128 \sqrt{\pi} \: t^{3/2}}
    \left[
    \erfc \left( \frac{y}{\sqrt{t}} \right)
    - \erfc \left( \frac{y}{\sqrt{1-t}} \right)
    \right]^2
    \:,
    \quad
    Q_{3;\Dt''}(y,0) = 0
    \:.
\end{equation}
The solution can be written as a double convolution with the heat kernel, which at $t=1$ takes the form,
\begin{equation}
    Q_{3;\Dt''}(y,1) = \int_{-\infty}^{\infty} \dd z \int_0^1 \dd t
    \:
    \frac{z \: \e^{-\frac{z^2}{t}}}{128 \sqrt{\pi} \: t^{3/2}}
    \left[
    \erfc \left( \frac{z}{\sqrt{t}} \right)
    - \erfc \left( \frac{z}{\sqrt{1-t}} \right)
    \right]^2
    \frac{\e^{-\frac{(y-z)^2}{1-t}}}{\sqrt{\pi(1-t)}}
    \:.
\end{equation}
The integrals over $z$ can be computed explicitly using the table~\cite{Owen:1980SM}. After multiple integrations by parts on the remaining time integral, we obtain for $y>0$,
\begin{multline}
    Q_{3;\Dt''}(y,1) =
    \frac{\e^{-y^2}}{64\pi}
    - \frac{1}{128} \erfc(y)
    - \frac{\sqrt{3}}{128 \pi} \erfc(y)
    - \frac{\e^{-2y^2}}{64 \pi} \erfc(y)
    + \frac{y \: \e^{-y^2}}{128 \sqrt{\pi}} \erfc(y)^2
    \\
    + \frac{1}{128} \erfc\left( \frac{y}{\sqrt{2}} \right)^2
    + \frac{\sqrt{3}}{128 \pi} \erfc \left( \sqrt{3} \: y \right)
    + \frac{1}{64} f(y)
    \:,
\end{multline}
where the function $f$ is given by
\begin{equation}
    \label{eq:FctF}
    f(y) = \int_0^1 \dd t \left[
       \frac{\e^{-\frac{y^2}{1-t^2}}}{\pi \sqrt{1-t^2}} \:
       \erf \left(
       \frac{t y}{\sqrt{(1-t^2)(1+2t)}}
       \right)
       - \frac{y(2+t)}{(1-t)(1+2t)^3/2 \pi^{3/2}}
       \e^{-\frac{(1+t)y^2}{(1-t)(1+2t)}}
    \right]
    \:.
\end{equation}
In particular, we have that
\begin{equation}
    f(0) = 0
    \:,
    \quad
    f'(0) = \frac{\sqrt{3}-1}{\pi^{3/2}}
    - \frac{1}{2\sqrt{\pi}}
    \:,
    \quad
    \int_0^\infty f(y) \dd y
    = \frac{1}{4\sqrt{\pi}} -\frac{3(\sqrt{3}-1)}{2 \pi^{3/2}}
    \:.
\end{equation}
These values will be useful to compute the fourth cumulant $\kappa_4$, and to check the guess~\eqref{eq:ConjPsiSM} of the cumulant generating function.

Additionally, after a few final integrations by parts, we can show that the function $f$ satisfies
\begin{equation}
    \label{eq:RHSforf}
    f''(y) + 2 y f'(t) =
    - \frac{2}{\pi^{3/2}} \partial_y \int_0^1 \frac{\dd t}{\sqrt{1+2t}}
    \e^{- \frac{(1+t) y^2}{(1-t)(1+2t)}}
    \:.
\end{equation}

\subsubsection{The term in \texorpdfstring{$\Dt' \sgt'$}{D sigma}}

The function $Q_{3;\Dt',\sgt'}$ also obeys a heat equation with a source term which is rather cumbersome, so we do not write it explicitly here. This source term is the sum of several expressions, some of which are only functions of the scaling variable $y/\sqrt{t}$. These parts can be solved by looking for a solution which is a function of this variable only. The heat equation then reduces to an ordinary differential equation, which can be solved. From this procedure, we obtain that,
\begin{multline}
    Q_{3;\Dt',\sgt'} =
    -\frac{y \: e^{-\frac{y^2}{t}}}{64 \sqrt{\pi t}}\erf\left(\frac{y}{\sqrt{t}}\right)^2
    -\frac{e^{-\frac{y^2}{t}} }{64 \pi\sqrt{t}} \erf\left(\frac{y}{\sqrt{t}}\right)
    -\frac{e^{-\frac{2 y^2}{t}}}{64\pi} \erf\left(\frac{y}{\sqrt{t}}\right)
    -\frac{1}{384} \erf\left(\frac{y}{\sqrt{t}}\right)^3
    +\frac{1}{384} \erf\left(\frac{y}{\sqrt{t}}\right)
    \\
    +\frac{y e^{-\frac{y^2}{t}}}{64 \sqrt{\pi t}}
    + \tilde{Q}_{3;\Dt',\sgt'}(y,t)
    \:,
\end{multline}
where $\tilde{Q}_{3;\Dt',\sgt'}$ is solution of
\begin{equation}
    \partial_t \tilde{Q}_{3;\Dt',\sgt'} (y,t)
    - \frac{1}{4} \partial_y^2 \tilde{Q}_{3;\Dt',\sgt'}(y,t)
    =
    \frac{y \sqrt{1-t}}{32 \pi^{3/2} t^2} \e^{-\frac{(2-t)y^2}{t(1-t)}}
    - \frac{t-2y^2}{64 \pi t^2} \e^{-\frac{2y^2}{t}}
    \erf \left( \frac{y}{\sqrt{1-t}} \right)
    \:,
    \quad
    Q_{3;\Dt',\sgt'}(y,0) = 0
    \:.
\end{equation}
The solution at $t=1$ can again be written in terms of a double convolution with the heat kernel. Computing first the spatial integrals using the table~\cite{Owen:1980SM}, and then the time integral (using also integration by parts), we get
\begin{equation}
    \tilde{Q}_{3;\Dt',\sgt'}(y,t)
    = \frac{\e^{-y^2}}{64 \pi} - \frac{\e^{-2y^2}}{64 \pi}
    \:,
    \quad \text{for} \quad
    y > 0
    \:.
\end{equation}

\subsubsection{The term in \texorpdfstring{$(\Dt')^2$}{D square}}

The last function to determine, $Q_{3;(\Dt')^2}(y,t)$, also obeys a heat equation, with a cumbersome source term which we do not reproduce here.
% We again isolate the terms which are only function of $y/\sqrt{t}$ and solve for these source terms by reducing the PDE to an ODE. This gives,
% \begin{multline}
%     Q_{3;(\Dt')^2}(y,t) =
%     \frac{y^2 e^{-\frac{y^2}{t}}}{64 \pi t^{3/2}}\erf\left(\frac{y}{\sqrt{t}}\right)
%     +\frac{y^3 e^{-\frac{y^2}{t}}}{128\sqrt{\pi}t^{3/2}} \erf\left(\frac{y}{\sqrt{t}}\right)^2
%     +\frac{y^2 e^{-\frac{2 y^2}{t}}}{64\pi t} \erf\left(\frac{y}{\sqrt{t}}\right)
%     -\frac{5 y e^{-\frac{y^2}{t}} }{256 \sqrt{\pi t}} \erf\left(\frac{y}{\sqrt{t}}\right)^2
%     \\
%     -\frac{e^{-\frac{y^2}{t}} }{128 \pi \sqrt{t}} \erf\left(\frac{y}{\sqrt{t}}\right)
%     -\frac{e^{-\frac{2 y^2}{t}}}{32 \pi} \erf\left(\frac{y}{\sqrt{t}}\right)
%     -\frac{3 \sqrt{3}}{256 \pi } \erf\left(\frac{y}{\sqrt{t}}\right)
%     +\frac{3 \sqrt{3}}{256 \pi} \erf\left(\frac{\sqrt{3} \: y}{\sqrt{t}}\right)
%     \\
%     +\frac{y e^{-\frac{y^2}{t}}}{128 \sqrt{\pi } \sqrt{t}}
%     +\frac{y e^{-\frac{3 y^2}{t}}}{128 \pi^{3/2} \sqrt{t}}
%     +\frac{y e^{-\frac{2 y^2}{t}}}{64 \pi ^{3/2} t}
%   + \tilde{Q}_{3;(\Dt')^2}(y,t)
%   \:,
% \end{multline}
% with $\tilde{Q}_{3;(\Dt')^2}$ that obeys a slightly less cumbersome source term, with $\tilde{Q}_{3;(\Dt')^2}(y,0) = 0$. We again express the solution at final time in terms of a double convolution with the heat kernel, and compute the spatial integral using~\cite{Owen:1980}.
In principle, one should be able to obtain explicitly $Q_{3;(\Dt')^2}$ using the same procedure as above, but that would require lengthy computations. Instead, we use the fact that the general result~\eqref{eq:q3} for any $\Dt(\rbt)$ and $\sgt(\rbt)$ should reduce to the known expressions which are available for specific choices of $\Dt$ and $\sgt$. Since we are looking for the term in $(\Dt')^2$, we cannot directly use the known results for the SEP, for which $\Dt' = 0$. However, we can use the known results for the random average process, given in~\cite{Grabsch:2022SM,Grabsch:2023SM}, corresponding to~\cite{Krug:2000SM,Kundu:2016SM}
\begin{equation}
    \label{eq:DSigmaRAP}
    \Dt(\rbt) = \frac{\mu_1}{2\rbt^2}
    \:,
    \quad
    \sgt(\rbt) = \frac{\mu_1 \mu_2}{\mu_1 - \mu_2} \frac{1}{\rbt}
    \:,
\end{equation}
with $\mu_1 > \mu_2 > 0$. Indeed, in this case, the profile $\qt_3(x,1)$ for the current in the RAP can be deduced from the profile $q_3(x,1)$ for the tracer in the dual model, which is the Kipnis-Marchioro-Presutti model, whose profiles are given in~\cite{Grabsch:2023SM}. The duality transformation~\eqref{eq:MappingDens} gives the profile
\begin{multline}
    \label{eq:q3RAP}
    \qt_{3,\mathrm{RAP}}(x,1) =
    \frac{1}{8} \erf(y)^2 \erfc(y)
    +\frac{1}{4} \erf(y) \erfc(y)
    -\frac{y^2 e^{-2 y^2}}{2 \pi } \erfc(y)
    -\frac{y^2 e^{-y^2}}{2 \pi } \erfc(y)
    -\frac{3 y \: e^{-y^2}}{4 \sqrt{\pi}} \erfc(y)^2
    \\
    +\frac{y \: e^{-y^2}}{\sqrt{\pi }}\erfc(y)
    +\frac{3 e^{-2 y^2}}{4\pi } \erfc(y)
    +\frac{3 e^{-y^2}}{4 \pi } \erfc(y)
    +\frac{y^3 e^{-y^2}}{4 \sqrt{\pi }} \erfc(y)^2
    +\frac{1}{8} \erfc\left(\frac{y}{\sqrt{2}}\right)^2
    \\
    +\frac{1}{2 \pi} \erfc(y)
    -\frac{7}{24}\erfc(y)
    +\frac{y \: e^{-3 y^2}}{4 \pi^{3/2}}
    +\frac{y \: e^{-2 y^2}}{2 \pi^{3/2}}
    +\frac{y \: e^{-y^2}}{4 \pi^{3/2}}
    -\frac{e^{-2 y^2}}{\pi }
    -\frac{e^{-y^2}}{2 \pi}
    \:.
\end{multline}
Imposing that the function $\qt_3$~\eqref{eq:q3} reduces to~\eqref{eq:q3RAP} for the specific choice~\eqref{eq:DSigmaRAP}, we find that
\begin{multline}
    Q_{3;(\Dt')^2} =
    -\frac{y^2 \: e^{-2 y^2}}{64 \pi }  \erfc(y)
    -\frac{y^2 e^{-y^2}}{64 \pi } \erfc(y)
    -\frac{5 y \: e^{-y^2} }{256 \sqrt{\pi }} \erfc(y)^2
    +\frac{y \: e^{-y^2}}{64 \sqrt{\pi }} \erfc(y)
    +\frac{e^{-2 y^2}}{32 \pi } \erfc(y)
    +\frac{e^{-y^2}}{128 \pi } \erfc(y)
    \\
    +\frac{y^3 e^{-y^2}}{128 \sqrt{\pi }} \erfc(y)^2
    -\frac{3}{256} \erfc\left(\frac{y}{\sqrt{2}}\right)^2
    +\frac{3\sqrt{3}}{256 \pi } \erfc(y)
    +\frac{1}{64 \pi }\erfc(y)
    +\frac{3}{256} \erfc(y)
    -\frac{3 \sqrt{3}}{256 \pi } \erfc\left(\sqrt{3} y\right)
    +\frac{y \: e^{-3 y^2}}{128 \pi ^{3/2}}
    \\
    +\frac{y \: e^{-2 y^2}}{64 \pi ^{3/2}}
    +\frac{y \: e^{-y^2}}{128 \pi ^{3/2}}
    -\frac{e^{-2 y^2}}{64 \pi}
    -\frac{5 e^{-y^2}}{128 \pi }
    -\frac{3}{128}f(y)
    \:,
\end{multline}
where $f$ is the function defined by the integral representation~\eqref{eq:FctF} above.

\section{Cumulants and profiles}

\subsection{Cumulants}

\subsubsection{Derivation of the general expression}

The cumulants are obtained by expanding the generating function $\hat\psi$ in powers of $\lambda$,
\begin{equation}
    \hat\psi = \kappa_2 \frac{\lambda^2}{2}
    + \kappa_4 \frac{\lambda^4}{4!}
    + \cdots
    \:.
\end{equation}
Expanding~\eqref{eq:CumulGenFctFromQ} in powers of $\lambda$, using~\eqref{eq:Exppqlambda}, we obtain,
\begin{equation}
    \kappa_2 = 2 \int_0^\infty \qt_1(x,1) \dd x
    \:,
    \quad
    \kappa_4 = 12 \int_0^\infty \qt_3(x,1) \dd x
    \:.
\end{equation}
Using the expressions of Section~\ref{sec:PertSol}, we obtain,
\begin{equation}
    \kappa_2 = \frac{\sgt(\rbt)}{\sqrt{\pi \Dt(\rbt)}}
    \:,
\end{equation}
\begin{multline}
    \kappa_4 = \frac{\sgt (\rbt) \sgt'(\rbt ) \left(\Dt(\rbt )\sgt'(\rbt)-\sgt(\rbt) \Dt'(\rbt)\right)}{4 \sqrt{\pi } \Dt(\rbt)^{7/2}}
    -\frac{3 \sgt(\rbt)^3 \left(\Dt'(\rbt)^2-2 \Dt(\rbt)\Dt''(\rbt )\right)}{8 \pi ^{3/2} \Dt(\rbt)^{9/2}}
    +\frac{3 \sgt(\rbt)^3 \left(\Dt'(\rbt)^2-\Dt(\rbt) \Dt''(\rbt)\right)}{8 \sqrt{\pi } \Dt(\rbt)^{9/2}}
    \\
    +\frac{\left(3 \sqrt{2}-4\right) \sgt (\rbt)^2 \sgt''(\rbt)}{8 \sqrt{\pi} \Dt(\rbt)^{5/2}}
    +\frac{3 \left(\sqrt{2} \pi -2 \sqrt{3}\right) \sgt (\rbt)^3 \left(2 \Dt(\rbt ) \Dt''(\rbt )-3 \Dt'(\rbt)^2\right)}{16 \pi ^{3/2} \Dt(\rbt )^{9/2}}
    \:.
\end{multline}
In this form, the cumulants are expressed in terms of the transport coefficients $\Dt$ and $\sgt$ of the dual model, and the mean density $\rbt$ in this model. It thus corresponds to the cumulants of the integrated current $\tilde{Q}_T$. To obtain the cumulants of the tracer, we rewrite these expressions in terms of the coefficients $D$ and $\sigma$ of the original model using~\eqref{eq:MappingTrCoefsSM}, and the mean density $\rb$ using~\eqref{eq:TrMeanDens}. This gives,
\begin{equation}
    \kappa_2 = \frac{\sigma(\rb)}{\rb^2 \sqrt{\pi D(\rb)}}
    \:,
\end{equation}
\begin{multline}
    \label{eq:Kappa4SM}
    \kappa_4 =
    \frac{3 \sigma (\rb )^3 \left(\rb  D'(\rb )+D(\rb )\right)}
    {\pi ^{3/2} \rb^6 D(\rb )^{7/2}}
   -\frac{\sigma (\rb )\left(
   \sigma (\rb )\sigma'(\rb ) \left(\rb  D'(\rb )+4 D(\rb )\right)
   + 2 \sigma (\rb )^2 D'(\rb )-\rb  D(\rb ) \sigma'(\rb )^2
   \right)}
   {4 \sqrt{\pi} \rb^5 D(\rb )^{7/2}}
   \\
   +\frac{3 \sigma (\rb )^3  \left(D'(\rb )^2-D(\rb ) D''(\rb )\right)}
   {8 \sqrt{\pi } \rb^4 D(\rb)^{9/2}}
   +\frac{3 \sigma (\rb)^3 \left(2 D(\rb ) D''(\rb )-D'(\rb )^2\right)}
   {8 \pi ^{3/2} \rb ^4 D(\rb )^{9/2}}
   +\frac{\left(3 \sqrt{2}-4\right) \sigma (\rb )^2 \sigma ''(\rb )}
   {8 \sqrt{\pi } \rb ^4 D(\rb )^{5/2}}
   \\
   + \frac{3 \left(\sqrt{2} \pi -2 \sqrt{3}\right) \sigma (\rb)^3 \left(2 D(\rb ) D''(\rb )-3 D'(\rb )^2\right)}
   {16 \pi^{3/2} \rb ^4 D(\rb)^{9/2}}
    \:.
\end{multline}
This is the result~(9) announced in the main text.

\subsubsection{Recovering known results in specific cases}

From our general expression of $\kappa_4$~\eqref{eq:Kappa4SM}, we easily check that we properly reproduce all previously known expressions of the fourth cumulant of the position of a tracer for specific models. We list below a few models, and the corresponding expression of $\kappa_4$ which coincides with those given in~\cite{Rizkallah:2022SM} (see~\cite{Krapivsky:2014SM} for the first determination of $\kappa_4$ for the SEP)
\[
\renewcommand{\arraystretch}{2}
\begin{array}{l*3{|>{\displaystyle}c}}
\text{Model} & D(\rho)  & \sigma(\rho) & \kappa_4
\\ \hline
\text{Simple exclusion process} & D_0 & 2 D_0 \rho(1-\rho)
& \frac{2\sqrt{D_0}(1-\rb)}{\pi^{3/2} \rb^3}
    \left(
        12 (1-\rb)^2 - \pi (3 - 3 (4 - \sqrt{2})\rb + (8-3\sqrt{2}) \rb^2)
    \right)\\[0.1cm]
\text{Random average process} & \frac{\mu_1}{2 \rho^2} & \frac{1}{\rho} \frac{ \mu_1 \mu_2}{\mu_1 - \mu_2}
& \frac{6 \mu_2^3}{(\mu_1-\mu_2)^3 \rb^4} \sqrt{\frac{\mu_1}{\pi}}\\[0.3cm]
\text{Kipnis-Marchioro-Presutti} & D_0 & \sigma_0 \rho^2
& \sigma_0^3 \frac{12+ \pi(3 \sqrt{2}-8)}{4 D_0^{5/2} \pi^{3/2}}\\[0.3cm]
\text{Hard Brownian particles} & D_0 & 2D_0 \rho
& \frac{6\sqrt{D_0}(4-\pi)}{\pi^{3/2} \rb^3}\\[0.1cm]
\text{Hard rod gas} & \frac{D_0}{(1-\ell \rho)^2} & 2 D_0 \rho
& \frac{6\sqrt{D_0}(4-\pi)(1-\ell \rb)^3}{\pi^{3/2} \rb^3}\\[0.3cm]
\text{Double exclusion process} & \frac{D_0}{(1-\rho)^2} & \frac{2 D_0 \rho(1-2\rho)}{1-\rho}
& \frac{2\sqrt{D_0}(1-2\rb)}{\pi^{3/2} \rb^3}
    \left( 12 (1-2 \rb )^2-\pi  ((23-6 \sqrt{2}) \rb^2-3
   (6-\sqrt{2}) \rb +3) \right)
\end{array}
\]

We briefly describe the models listed in this table. In the simple exclusion process (SEP), each particle can hop onto a neighbouring site if it is empty. The random average process describes particles on a continuous line which can hop to a random fraction of the distance to a neighbouring particle. The Kipnis-Marchioro-Presutti describes a lattice in which each site hosts a continuous variable (a mass). At random times the total mass of two neighbouring sites is randomly redistributed between them. The hard Brownian particles is a model of hardcore particles performing Brownian motion, with the constraint that they remain in the same order. The hard rod gas is similar, but the particles have a finite length $\ell$ and cannot overlap. Finally, the double exclusion process is similar to the SEP, but the particles occupy two neighbouring sites and can only hop by one unit if the site in the randomly chosen direction is empty.

\subsection{Correlation profiles}

The correlation profiles can be obtained from the MFT~\cite{Poncet:2021SM} of the dual model using the mapping~\eqref{eq:MappingDens},
\begin{equation}
    w(x,T)
    \equiv
    \frac{\moy{ \rho(X_T+x,T) \e^{\lambda X_T}}}
    {\moy{\e^{\lambda X_T}}}
    = \frac{\moy{ \frac{1}{\rt(Y_x[\rt],T)} \e^{-\lambda \tilde{Q}_T}}}
    {\moy{\e^{-\lambda \tilde{Q}_T}}}
    \:,
    \quad
    Y_x[\rt] = \int_0^x \rt(y',T) \dd y'
    \:.
\end{equation}
Using the formalism of Section~\ref{sec:MFT}, we can write this profile as (after rescaling space by $\sqrt{T}$ and time by $T$),
\begin{equation}
    w(x \sqrt{T},T)
    = \frac{
    \displaystyle
    \int \mathcal{D} \tilde\rho(x,t) \mathcal{D} \tilde{H}(x,t)
    \int \mathcal{D} \tilde\rho(x,0) \frac{1}{\rt(Y_x[\rt],1)}
    \e^{- \sqrt{T}(\lambda Q[\tilde\rho] + S[\tilde{\rho},\tilde{H}] + F[\tilde{\rho}(x,0)])}
    }{
    \displaystyle
    \int \mathcal{D} \tilde\rho(x,t) \mathcal{D} \tilde{H}(x,t)
    \int \mathcal{D} \tilde\rho(x,0) \:
    \e^{- \sqrt{T}(\lambda Q[\tilde\rho] + S[\tilde{\rho},\tilde{H}] + F[\tilde{\rho}(x,0)])}
    }
    \:.
\end{equation}
These integrals can again be evaluated by a saddle point method for large $T$. The prefactor in the numerator does not change the saddle point, and thus,
\begin{equation}
    w(z \sqrt{T},T) \underset{T \to \infty}{\simeq}
    \frac{1}{\qt(Y_z[\qt],1)}
    \equiv \Phi(z)
    \:,
    \quad
    Y_z[\qt] = \int_0^z \qt(y,1) \dd y
    \:.
\end{equation}
Using the above expressions for $\qt(y,1)$ at lowest orders in $\lambda$, we can perform this mapping explicitly to get
\begin{equation}
    \Phi(z) = \rb + \lambda \Phi_1(z)
    + \lambda^2 \Phi_2(z)
    + \lambda^3 \Phi_3(z)
    + \mathcal{O}(\lambda^4)
    \:,
\end{equation}
where, for $z>0$,
\begin{equation}
    \label{eq:Phi1}
    \Phi_1(z) = \frac{\sigma(\rb)}{4 \rb D(\rb)}
    \erfc \left(y  = \frac{z}{2\sqrt{D(\rb)}} \right)
    \:,
\end{equation}
\begin{multline}
    \label{eq:Phi2}
   \Phi_2(z) =
   \frac{\sigma (\rb) \sigma '(\rb)}{16 \rb ^2 D(\rb )^2}
   \erfc(y)
   -\frac{\sigma (\rb)^2}{4 \pi  \rb ^3 D(\rb )^2}e^{-y^2}
   \\
   + \frac{\sigma (\rb )^2 D'(\rb)}{32 \pi  \rb^2 D(\rb)^3}
   \left[
   \pi  \erf(y) \erfc(y)+2 \sqrt{\pi } e^{-y^2} y \erfc(y)-2 (e^{-2 y^2}+e^{-y^2})
   \right]
   \:,
\end{multline}
\begin{multline}
    \label{eq:Phi3}
    \Phi_3(z) =
   \frac{\sigma(\rb)^3}{16\pi ^{3/2} \rb^5 D(\rb)^3}
   \left(\sqrt{\pi} \text{erfc}(y)+2 e^{-y^2} y\right)
   +\frac{\sigma (\rb) \sigma'(\rb)^2}{96 \rb^3 D(\rb)^3}\text{erfc}(y)
   -\frac{\sigma (\rb)^2 \sigma '(\rb)}{96 \rb^4 D(\rb)^3}
   \left(\text{erfc}(y)+\frac{6 e^{-y^2}}{\pi }\right)
   \\
   +\frac{\sigma (\rb)^2\sigma ''(\rb)}{384 \rb^3 D(\rb)^3}
   \left(3 \text{erfc}\left(\frac{y}{\sqrt{2}}\right)^2-2 \text{erfc}(y)\right)
   -\frac{\sigma (\rb)^2 D'(\rb) \sigma'(\rb)}{192 \rb^3 D(\rb)^4}
   \left(-\frac{6 e^{-y^2} y \text{erfc}(y)}{\sqrt{\pi}}+3\text{erfc}(y)^2-\text{erfc}(y)+\frac{6 e^{-2 y^2}}{\pi }\right)
    \\
    +\frac{\sigma(\rb)^3 D'(\rb)}{192 \pi^{3/2} \rb^4 D(\rb)^4}
    \left(-12 \sqrt{\pi } e^{-y^2} y^2 \text{erfc}(y)+18 \sqrt{\pi } e^{-y^2}
   \text{erfc}(y)-\pi ^{3/2} \text{erfc}(y)+12 \sqrt{\pi } \text{erfc}(y)+12 e^{-2 y^2}
   y+12 e^{-y^2} y-12 \sqrt{\pi } e^{-y^2}\right)
   \\
   +\frac{\sigma (\rb)^3 D'(\rb)^2}{256 \rb^3 D(\rb)^5}
   \Bigg(-\frac{4 e^{-2 y^2} y^2 \text{erfc}(y)}{\pi }-
   \frac{4 e^{-y^2} y^2 \text{erfc}(y)}{\pi }
   +\frac{4 e^{-y^2} y \text{erfc}(y)}{\sqrt{\pi }}
   +\frac{6 e^{-y^2} \text{erfc}(y)}{\pi}
   +\frac{2 e^{-y^2} y^3 \text{erfc}(y)^2}{\sqrt{\pi }}
   -2 \text{erfc}(y)^2
   \\
   +\frac{4\text{erfc}(y)}{\pi }+\frac{2 e^{-3 y^2} y}{\pi ^{3/2}}+\frac{4 e^{-2 y^2} y}{\pi
   ^{3/2}}+\frac{2 e^{-y^2} y}{\pi ^{3/2}}-\frac{4 e^{-2 y^2}}{\pi }
   \Bigg)
   \\
   +\frac{\sigma (\rb)^3 \left(2 D(\rb) D''(\rb)-3 D'(\rb)^2\right)}{256 \rb ^3 D(\rb)^5}
   \left(-\frac{e^{-y^2}  y \text{erfc}(y)^2}{\sqrt{\pi }}-\frac{\sqrt{3} \text{erfc}(y)}{\pi}-\text{erfc}(y)+\text{erfc}\left(\frac{y}{\sqrt{2}}\right)^2
   +\frac{\sqrt{3}}{\pi }\text{erfc}\left(\sqrt{3} y\right)
   +\frac{2 e^{-y^2}}{3 \pi} +2 f(y)
   \right)
   \\
   -\frac{\sigma (\rb)^3 \left(D(\rb) D''(\rb)-3 D'(\rb)^2\right)}{384 \rb^3 D(\rb)^5}
   \left(-\frac{6 e^{-y^2} y \text{erfc}(y)^2}{\sqrt{\pi }} +\frac{6 e^{-2
   y^2} \text{erfc}(y)}{\pi }+\text{erfc}(y)^3-\frac{4 e^{-y^2}}{\pi }\right)
   \:,
\end{multline}
where $f$ is the function~\eqref{eq:FctF}. Finally, Eqs.~(\ref{eq:Phi1}-\ref{eq:Phi3}) give the expression of the bath-tracer correlation profiles up to order $3$.

\subsection{An equation for the profiles?}

In Refs.~\cite{Grabsch:2022SM,Grabsch:2023SM}, the derivation of the lowest orders $\Phi_n$ of the correlation profile for the SEP allowed for the determination of a simple closed equation satisfied by the profile $\Phi$ at all orders in $\lambda$. The starting point was the equation verified by the profiles of the SEP in the low density limit~\cite{Poncet:2021SM},
\begin{equation}
    \label{eq:BulkEqLowDens}
    \Phi''(z) + \frac{1}{2} (z + \xi) \Phi'(z) = 0
    \:,
    \quad
    \text{with}
    \quad
    \xi = \dt{\hat\psi}{\lambda}
    \:,
    \quad
    \text{for the SEP with } \rb \to 0
    \:.
\end{equation}
The approach used in Refs.~\cite{Grabsch:2022SM,Grabsch:2023SM} relied on three steps: (i) look for an equation with the same l.h.s., and compute the new r.h.s. using the expressions of $\Phi_n$ obtained by MFT; (ii) rewrite this r.h.s. in a closed form in terms of $\Phi$ only and finally (iii) infer the general structure from the lowest orders of this equation. This led to a guess for a simple closed equation, that was later proved from MFT~\cite{Mallick:2022SM}.

We investigate here, for arbitrary $D(\rho)$ and $\sigma(\rho)$, the possibility to write such an equation by following the same procedure. Since the SEP was for constant diffusion coefficient, we modify the l.h.s. of~\eqref{eq:BulkEqLowDens} to include $D(\rho)$ in the first term (which comes from the diffusion equation~\eqref{eq:MFT_qSM}). For step (i), we compute (for $z > 0$, the case $z<0$ can be deduced from the symmetry $z \to -z$, $\lambda \to -\lambda$),
\begin{multline}
    \label{eq:bulkEq0}
    \partial_z(D(\Phi) \Phi') + \frac{1}{2}(z + \xi) \Phi'(z)
    = \lambda^2 \frac{\sigma(\rb)^2 D'(\rb)}{32 \pi \rb^2 D(\rb)^3}e^{-y^2}
    -\lambda ^3 \frac{\sigma(\rb)^2 \sigma''(\rb)}{256 \pi \rb^3 D(\rb)^3}
    \left(\sqrt{2 \pi } e^{-\frac{y^2}{2}} y
   \erfc\left(\frac{y}{\sqrt{2}}\right)-2 e^{-y^2}\right)
   \\
   + \lambda ^3 \frac{\sigma (\rb)^3 \left(2 D(\rb) D''(\rb)-3 D'(\rb)^2\right)}
   {512 \rb^3 D(\rb)^5}
   \left(-\sqrt{\frac{2}{\pi }} \: y \: e^{-\frac{y^2}{2}}
   \erfc\left(\frac{y}{\sqrt{2}}\right)
   + f''(y)+2 y f'(y)\right)
   \\
   +
   \lambda ^3 \frac{\sigma (\rb)^3 D'(\rb)^2}{256 \pi ^{3/2} \rb ^3 D(\rb)^5}
   \left(2 \sqrt{\pi } e^{-y^2} y^2 \erfc(y)-3 \sqrt{\pi } e^{-y^2}
   \erfc(y)-2 y e^{-2 y^2} -2 y e^{-y^2}+2 \sqrt{\pi } e^{-y^2}\right)
   \\
   + \lambda^3 \frac{ \sigma (\rb)^3 D'(\rb)}{64\pi ^{3/2} \rb^4 D(\rb)^4}
   e^{-y^2} \left(\sqrt{\pi }-2 y\right)
   + \mathcal{O}(\lambda^4)
   \:,
\end{multline}
again with $y = \frac{z}{2\sqrt{D(\rb)}}$. A key step in~\cite{Grabsch:2022SM,Grabsch:2023SM}, in which only the term in $\sigma''(\rb)$ was present, was to notice that this term can be written in a closed form using
\begin{equation}
    \int_0^\infty \dd u \: \Phi''(-u) \Phi'(z+u)
    = -\lambda ^2  \frac{\sigma (\rb)^2}{64 \pi \rb ^2 D(\rb)^3}
   \left(\sqrt{2 \pi } e^{-\frac{y^2}{2}} y
   \erfc\left(\frac{y}{\sqrt{2}}\right)-2 e^{-y^2}\right)
   + \mathcal{O}(\lambda^3)
   \:.
\end{equation}
Here, we go further by noticing that several terms in~\eqref{eq:bulkEq0} can be written in terms of $\Phi'(z)$, so that
\begin{multline}
    \label{eq:BulkEqSM}
    \partial_z(D(\Phi)\partial_z \Phi)
    + \frac{1}{2}(z+\xi) \partial_z \Phi
    = \frac{\lambda \sigma''(\rb)}{4\rb} \int_0^\infty \Phi'(z+u)\Phi''(-u) \dd u
    \\
    + \left( \frac{\lambda  \sigma (\rb ) D'(\rb )}{8 \sqrt{\pi } \rb  D(\rb )^{3/2}}
    -\frac{\lambda ^2 \sigma (\rb ) D'(\rb ) \left(\rb  \sigma '(\rb )-2 \sigma (\rb)\right)}{32 \sqrt{\pi } \rb ^3 D(\rb )^{5/2}}
    +\frac{\lambda ^2 \sigma (\rb )^2 D'(\rb )^2}{64 \sqrt{\pi } \rb ^2 D(\rb)^{7/2}}
    \right) \Phi'(z)
    \\
    - \frac{\lambda ^3 \sigma (\rb )^3 \left(2 D(\rb ) D''(\rb )-3 D'(\rb )^2\right)}{512
   \rb ^3 D(\rb )^5}
   \left(
   y \e^{-\frac{y^2}{2}} \sqrt{\frac{2}{\pi}} \erfc \left( \frac{y}{\sqrt{2}} \right)
   +\frac{2}{\pi^{3/2}} \partial_y \int_0^1 \frac{\dd t}{\sqrt{1+2t}}
    \e^{- \frac{(1+t) y^2}{(1-t)(1+2t)}}
   \right)
   + \mathcal{O}(\lambda^4)
   \:,
\end{multline}
where we have used~\eqref{eq:RHSforf} to rewrite the last term. This is the equation~(11) given in the main text. The open challenge is to rewrite the last two terms in terms of $\Phi$ only. This result~\eqref{eq:BulkEqSM} is the first step towards obtaining a closed equation for $\Phi$ for arbitrary $D(\rho)$ and $\sigma(\rho)$.

\subsection{Models that can be mapped onto constant diffusion coefficient}

The last term in~\eqref{eq:BulkEqSM} vanishes if and only if
\begin{equation}
    2 D(\rho) D''(\rho) - 3 D'(\rho)^2 = 0
    \:.
\end{equation}
Solving this differential equation yields
\begin{equation}
    \label{eq:Dvanish}
    D(\rho) = \frac{1}{(a + b \rho)^2}
    \:,
\end{equation}
with $a$ and $b$ two integration constants. This corresponds to the class of diffusion coefficients that can be mapped onto constant $D(\rho)$ by an extension of the mapping~\eqref{eq:MappingDens}. The most general type of mapping between two single-file systems was discussed in~\cite{Rizkallah:2022SM}. If $b=0$, then~\eqref{eq:Dvanish} already constant. If $b \neq 0$, we define
\begin{equation}
    \tilde\rho(y,t) =c + \frac{1/b}{a + b \rho(x(y,t),t)}
    \:,
    \quad
    x(y,t) = \int_0^y \rt(y',t) \dd y' + X_t
    \:.
\end{equation}
Under this mapping, the single-file with $D(\rho)$~\eqref{eq:Dvanish} and any $\sigma(\rho)$ maps onto a new single-file system, with~\cite{Rizkallah:2022SM}
\begin{equation}
    \Dt(\rho) = 1
    \:,
    \quad
    \sgt(\rho) = b (c-\rho) \rho \: \sigma \left( \frac{1 + a b (c - \rho)}{b^2 (\rho-c)} \right)
    \:.
\end{equation}

\subsection{A conjecture for the cumulant generating function}

In the case of the SEP, with $D(\rho) = D_0$ and $\sigma(\rho) = 2 D_0 \rho(1-\rho)$, it was shown that (for $D_0 = \frac{1}{2}$)~\cite{Poncet:2021SM,Grabsch:2022SM,Grabsch:2023SM},
\begin{equation}
    \label{eq:BoundCondSEP}
    \frac{1-\Phi(0^-)}{1-\Phi(0^+)} = \e^\lambda
    \:,
    \quad
    \Phi'(0^\pm) = \mp \frac{1}{2 D_0} \frac{\hat\psi}{\e^{\pm \lambda}-1} \Phi(0^\pm)
    \:,
\end{equation}
where we have reintroduced a generic value of $D_0$. From the first equation, we deduce
\begin{equation}
    \label{eq:BoundP}
    \lambda = P(\Phi(0^+)) - P(\Phi(0^-))
    \:,
    \quad
    P(\rho) = \int^\rho \frac{2 r D(r)}{\sigma(r)} \dd r =
    - \ln(1-\rho) \quad \text{for the SEP.}
\end{equation}
Similarly, the equations~\eqref{eq:BoundCondSEP} can be combined to yield
\begin{equation}
    \frac{\Phi'(0^+)}{\Phi(0^+)(1-\Phi(0^+))}
    = \frac{\Phi'(0^-)}{\Phi(0^-)(1-\Phi(0^-))}
    \:,
\end{equation}
which can be written as
\begin{equation}
    \label{eq:BoundMu}
    \left. \partial_x \mu(\Phi) \right|_{0^+}
    = \left. \partial_x \mu(\Phi) \right|_{0^-}
    \:,
    \quad
    \mu(\rho) = \int^\rho \frac{2 D(r)}{\sigma(r)} \dd r
    \:.
\end{equation}
The two equations~(\ref{eq:BoundP},\ref{eq:BoundMu}) are the boundary conditions given in the main text. They have first been derived for the specific case of the SEP, and it has recently been shown that they hold for any single-file model with arbitrary $D(\rho)$ and $\sigma(\rho)$~\cite{Grabsch:2024SM}.

Although the two boundary conditions~(\ref{eq:BoundP},\ref{eq:BoundMu}) are very general, and take a compact physical form, they give less information than the three equations~\eqref{eq:BoundCondSEP}. In particular, Eqs.~(\ref{eq:BoundP},\ref{eq:BoundMu}) do not involve the cumulant generating function $\hat\psi$, while~\eqref{eq:BoundCondSEP} does. To fully generalise~\eqref{eq:BoundCondSEP} to any $D(\rho)$ and $\sigma(\rho)$, we need the last relation that relates $\hat\psi$ and $\Phi(0^\pm)$, $\Phi'(0^\pm)$. This relation was very useful in~\cite{Poncet:2021SM,Grabsch:2022SM,Grabsch:2023SM} as it allows to easily compute $\hat\psi$ from $\Phi$. To obtain this relation for any single-file system, let us start by rewriting~\eqref{eq:BoundCondSEP} as
\begin{equation}
    \hat\psi = - 2 D_0 \frac{\Phi'(0^+)}{\Phi(0^+)}(\e^\lambda - 1)
    = - 2 D_0 \frac{\Phi'(0^+)}{\Phi(0^+)(1-\Phi(0^+))}(\Phi(0^+) - \Phi(0^-))
    \:.
\end{equation}
We can rewrite this expression in terms of the chemical potential defined in~\eqref{eq:BoundMu}, which for the SEP yields
\begin{equation}
    \hat\psi
    = - 2 D_0 (\Phi(0^+) - \Phi(0^-)) \left. \partial_x \mu(\Phi) \right|_{0^+}
    \:.
\end{equation}
This form is consistent with the conjecture given in the main text, which reads
\begin{equation}
    \label{eq:ConjPsiSM}
    \hat\psi
    = - 2 \left. \partial_x \mu(\Phi) \right|_{0^+} \int_{\Phi(0^-)}^{\Phi(0^+)} D(r)\dd r
    \:.
\end{equation}

\bigskip

Furthermore, the mapping~\eqref{eq:MappingDens} between two different single-file systems impose a strong condition on $\hat\psi$: it must be the same function for a tracer in a single-file with $D$ and $\sigma$ and for the current in the dual single-file $\Dt$, $\sgt$ due to~\eqref{eq:RelXtQt}. Additionally, the profiles transform as~\eqref{eq:MappingDens},
\begin{equation}
    \tilde{\Phi}(y(x)) = \frac{1}{\Phi(x)}
    \:,
    \quad
    y(x) = \int_0^x \Phi(x') \dd x'
    \:,
\end{equation}
and the transport coefficients $D$, $\sigma$ as~\eqref{eq:MappingTrCoefsSM}. We can thus show that
\begin{equation}
    \partial_x \mu(\Phi)
    = \frac{1}{\tilde{\Phi}(y)} \partial_y \mu \left( \frac{1}{\tilde{\Phi}(y)} \right)
    = - \frac{\tilde{\Phi}'(y)}{\tilde{\Phi}(y)^3}
    \frac{2 D \left( \frac{1}{\tilde\Phi} \right)}{\sigma \left( \frac{1}{\tilde\Phi} \right)}
    = - \tilde{\Phi}'(y)
    \frac{2 \Dt( \tilde\Phi )}{\sigma( \tilde\Phi )}
    = - \partial_y \tilde{\mu}(\tilde{\Phi})
    \:,
\end{equation}
where $\tilde\mu(\rho) = \int^\rho \frac{2 \Dt}{\sgt}$ is the chemical potential in the dual model. Similarly, we have,
\begin{equation}
    \int_{\Phi(0^-)}^{\Phi(0^+)} D(r)\dd r
    = -\int_{1/\Phi(0^-)}^{1/\Phi(0^+)} \frac{\dd r}{r^2}
    D \left( \frac{1}{r} \right)
    = -\int_{\tilde\Phi(0^-)}^{\tilde\Phi(0^+)} \Dt(r) \dd r
    \:,
\end{equation}
so that Eq.~\eqref{eq:ConjPsiSM} is invariant under the duality mapping, as it should.

\bigskip

Finally, from our explicit results on $\kappa_4$~\eqref{eq:Kappa4SM} and the profiles~(\ref{eq:Phi1}-\ref{eq:Phi3}), we check that~\eqref{eq:ConjPsiSM} holds, at least up to order $4$ in $\lambda$ included.

\bigskip

All these arguments strongly support the validity of the conjecture~\eqref{eq:ConjPsiSM} for arbitrary $D(\rho)$ and $\sigma(\rho)$, at all orders in $\lambda$.

\bibliographystyle{apsrev4-1}
% \bibliography{bibSF}


\end{document}